# Embracing a mechanized formalization gap

Pragmatic software system verification (Extended version)

Antal Spector-Zabusky
Computer and Information Science
University of Pennsylvania
Philadelphia, PA, USA
antals@cis.upenn.edu

Yao Li
Computer and Information Science
University of Pennsylvania
Philadelphia, PA, USA
liyao@cis.upenn.edu

Joachim Breitner
DFINITY Foundation
Germany
joachim@dfinity.org

Stephanie Weirich
Computer and Information Science
University of Pennsylvania
Philadelphia, PA, USA
sweirich@cis.upenn.edu

**Abstract**

If a code base is so big and complicated that complete mechanical verification is intractable, can we still apply and benefit from verification methods? We show that by allowing a deliberate *mechanized formalization gap* we can shrink and simplify the model until it is manageable, while still retaining a meaningful, declaratively documented connection to the original, unmodified source code. Concretely, we translate core parts of the Haskell compiler GHC into Coq, using hs-to-coq, and verify invariants related to the use of term variables.

## 1   Introduction

Why don't we use proof assistants to reason about *existing* industrial-strength software systems?

Consider the Glasgow Haskell Compiler (GHC). It would be no surprise to learn that GHC is both very *big* and very *complicated*. Indeed, a fresh copy of GHC 8.4.3 contains 464 Haskell modules and 182 174 lines of code[1]. So, while *complete* verification of GHC is out of the question because of its scale and complexity, we don't have to prove everything about the entire system. Even if we only reason about *some* of the properties of *part* of the code base, we can still see benefits from mechanical reasoning, such as bug discovery, checked and explicit documentation of invariants, and a deeper understanding of the code.

In this paper, we demonstrate this approach via a case study of *partial* verification of GHC. The portion that we choose to focus on is GHC's intermediate language, *Core*, and the invariants related to the representation of term variables in that language. In particular, we analyze two selected operations that work with Core terms and prove that they preserve those invariants.

To start, we need a version of GHC's source code that allows for mechanized reasoning and that closely models the parts of GHC that we wish to reason about. The hs-to-coq tool [49] can be used to provide a shallow embedding of Haskell into Gallina, the language of the Coq proof assistant [15]. In prior work, Breitner et al. [11] used this tool to show that the Data.Set and Data.IntSet modules from the Haskell containers library are correct implementations of finite set data structures. In this work, we develop new methodology and extend the tool so that we can apply it to a substantial part of GHC.

In completing this work, we faced three main challenges. First, a software system on the scale of GHC includes Haskell features that have not yet been encountered in prior use cases of hs-to-coq. We needed novel ways to *reconcile* the discrepancies between the two languages. Second, because GHC is *big*, we do not want our embedding to start at the roots of the dependency hierarchy. We needed a way to *select* the code in the middle of the dependency graph that we care about, while abstracting the rest. Finally, because GHC is *complicated*, we needed a way to *simplify* and *refine* details of its implementation, focusing our attention on aspects of the system that we want to reason about while ignoring others.

We addressed these challenges through careful use of hs-to-coq's *edit files*, and by extending the edit language. *Edits* are instructions that control hs-to-coq's translation from Haskell into Gallina. Previous work used edits to bridge semantic differences between the two languages and make small-scale modifications to the translation in flight. In this verification effort, we increased the expressiveness of the edit language as one part of developing techniques for managing the complexities of systems at scale.

In particular, our work makes the following contributions:

---

[1]Nonblank, noncomment lines of code, calculated using the cloc tool, available from https://github.com/AlDanial/cloc.

- We construct a Gallina model of a subset of GHC, derived from the original Haskell source code. Although this model simplifies the representation of the Core intermediate language and its operations, it includes enough details to justify mechanical reasoning.
- We develop a formal specification of two Core invariants related to term variable binding, well-scopedness and the correct use of join points, demonstrating that this model is suitable for analysis (Section 3). These invariants are important to the correctness of the compiler and are drawn from comments in the GHC source code.
- We demonstrate some of the benefits of partial mechanical verification. When reasoning about substitution, we found that some of the comments specifying the required invariants were incorrect. When reasoning about `exitify`, we encountered a bug that we have repaired in GHC. We prove that our new version preserves both invariants (Section 4).
- We extend hs-to-coq with new forms of edits that allow for extensive reconciliation, selection, and simplification of the translated Gallina code (Section 5). We discuss in detail how we use those edits to adapt the GHC source code for reasoning (Section 6).

An alternative approach to reasoning about GHC would be to develop a Gallina model of the relevant parts of the system by hand. However, although our hs-to-coq translation simplifies some aspects of the implementation, it is unlikely that a hand-developed model would be as faithful of a representation. Our translation targets around 12% of GHC's source code, and produces over 18 000 lines of Gallina definitions. Our model is rich, detailed, and corresponds closely to the actual implementation.

All of our work is available online under the open-source MIT license, including the Core language model, the edits required for its creation, and the proofs of its properties.[2]

## 2  Case study: GHC

The case study that we have selected for this project is a portion of the Glasgow Haskell Compiler (GHC), version 8.4.3.[3] Specifically, GHC performs much of its optimization work on an intermediate language called Core; we selected as our target the goal of reasoning about the usage of variables during these Core-to-Core passes. This intermediate language is the target of GHC's type inference and desugarer, as shown in Figure 1. (This figure only provides detail for the first part of the GHC pipeline. The step marked "Rest of pipeline" includes many compilation steps for code generation, etc.)

Why does GHC make a good case study? First, it is one of the largest open source Haskell projects available. As a result, it is a realistic example of a complex system and a suitable challenge for our methodology. At the same time, this challenge is self-contained. Because GHC is a bootstrapping compiler, it relies only on a small number of external libraries.

Second, GHC is a mature project. The first prototype was built thirty years ago [29]. The current design of the Core language dates from the mid-2000s [50], but is still the target of significant revisions [10, 40]. Although the compiler has been around for a long time, it is under active development by a large, distributed team of contributors. The code itself is well documented, both internally (there are over 100 000 lines of comments) and externally (there is a wiki documenting the compiler,[4] and the source repository contains a guide to the design of Core [19]).

Third, GHC is an industrial-strength compiler for a real-world programming language and it is written with a heavy focus on performance. As as consequence, the correctness of GHC's optimizations for Core depends on several representation invariants of the Core language, and the code that must maintain these invariants is subtle and designed for performance [44]. Although the invariants are easy to specify (see Section 3), they cause difficulties for property-based testing [43]. And even though some parts of the implementation have been in GHC for over twenty years, we know of no attempt to mechanically reason about these invariants.

### 2.1  The Core AST in Haskell and Gallina

The Haskell version of the Core AST is shown on the left side of Figure 2.[5] This language is based on an explicitly-typed variant of System F called System FC [19, 50]. It includes variables (`Var`), constant literals (`Lit`), function applications (`App`), lambda abstractions (`Lam`), potentially-recursive let bindings (`Let`), and case expressions (`Case`). The `Tick` constructor marks profiling information and the remaining three data constructors carry information related to Core's type system.

The Core intermediate language uses a named representation for variables. GHC developers have found that working with concrete variable names, even though they require freshening to avoid capture, is the most efficient representation [44]. The `Expr` data type is parameterized by the type of bound variables, as can be seen in the `Lam` case. For the part of the code that we consider in this paper, this type is called `Var`; other parts of the compiler, which we do not interact with in this work, use a different type for variable bindings.

The `Expr` data type is already rather succinct in GHC; the desugarer converts the much larger source Haskell AST[6] into Core by elaborating the many forms of syntactic sugar found in Haskell. One source of brevity is the reuse of the `Lam`

---

[2]https://github.com/antalsz/hs-to-coq
[3]Available from https://www.haskell.org/ghc/
[4]https://gitlab.haskell.org/ghc/ghc/wikis/commentary/compiler/
[5]From module `CoreSyn` in the GHC implementation. This and all other code samples in the paper may be reformatted, have module names or comments removed, or similar for greater clarity.
[6]The `HsExpr` data type has 40 data constructors.

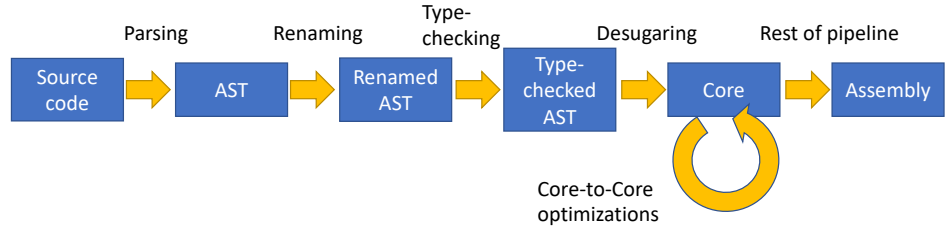

**Figure 1.** GHC compilation pipeline

```
data Expr b                              Inductive Expr b : Type
  = Var Id                               := | Mk_Var : Id -> Expr b
  | Lit Literal                             | Lit : Literal -> Expr b
  | App (Expr b) (Arg b)                    | App : (Expr b) -> (Expr b) -> Expr b
  | Lam b (Expr b)                          | Lam : b -> (Expr b) -> Expr b
  | Let (Bind b) (Expr b)                   | Let : (Bind b) -> (Expr b) -> Expr b
  | Case (Expr b) b Type [Alt b]            | Case : (Expr b) -> b -> Type_
  | Cast (Expr b) Coercion                    -> list ((fun b_ => (AltCon * list b_ * Expr b_)) b)
  | Tick (Tickish Id) (Expr b)                -> Expr b
  | Type Type                               | Cast : (Expr b) -> Coercion -> Expr b
  | Coercion Coercion                       | Mk_Type : Type_ -> Expr b
  deriving Data                             | Mk_Coercion : Coercion -> Expr b
data Bind b                              with Bind b : Type
  = NonRec b (Expr b)                    := | NonRec : b -> (Expr b) -> Bind b
  | Rec [(b, (Expr b))]                     | Rec : list (b * (Expr b)) -> Bind b.
  deriving Data
type Arg b = Expr b                      Definition Arg := Expr.
type Alt b = (AltCon, [b], Expr b)       Definition Alt := fun b_ => (AltCon * list b_ * Expr b_).
type Id = Var                            Definition Id := Var.
type CoreBndr = Var                      Definition CoreBndr := Var.
type CoreExpr = Expr CoreBndr            Definition CoreExpr := (Expr CoreBndr).
type CoreBind = Bind CoreBndr            Definition CoreBind := (Bind CoreBndr).
type CoreProgram = [CoreBind]            Definition CoreProgram := (list CoreBind).
```

**Figure 2.** Haskell (left) and hs-to-coq generated Gallina (right) versions of the Core AST.

constructor for abstraction over terms, types, and coercions. Similarly, GHC uses the Var type to represent term, type, and coercion variables. When used as term variables, as in the Var case of Expr, variables are called identifiers and are referred to by the type synonym Id.

The Gallina version of the Core AST (right side of Figure 2) renames the Var, Type and Coercion data constructors, because Coq uses only a single namespace for types and values. Furthermore, because the Type type constructor is a keyword in Coq, this name becomes Type_ in Gallina. The translation also removes the Tick constructor, as we do not wish to reason about profiling.

### 2.2 Core invariants and operations

Our case study focuses on syntactic invariants related to variable binding that are described in the comments of GHC and are validated by CoreLint, a developer-mode pass that checks that the output of each Core-to-Core pass is well-typed. The GHC developers report that the use of CoreLint has played a crucial role in eliminating many tricky bugs [39].

In particular, we consider the following two properties, discussed in more detail in Section 3:

- **The *well-scopedness* invariant** ensures that all local variables are in the scope of a matching binder. This fundamental property is a prerequisite for terms to represent lambda-calculus expressions and violating it is a source of subtle bugs.
- **The *join point invariant*** describes where and how *join points* (local jump targets) may be declared and called. This invariant is interesting because join points are innovative and a relatively new addition to GHC [40].

Our case study also focuses on two operations.

**Substitution:** The substExpr operation performs capture-avoiding substitution for term, type and coercion variables in the Core AST.

**Exitification:** The exitify optimization transforms code so that the compiler has more opportunities to inline definitions. In particular, it moves expressions on the exit path (identified by join points) out of recursive loops so that they may be inlined by the simplifier.

In particular, we prove that substitution preserves well-scopedness and that exitification preserves both invariants. We discuss these results in more detail in Section 4.

We selected these operations for our case study because of familiarity (one co-author is the developer of the exitify optimization), because they do interesting things to terms without relying on types, and because they modify the binding structure of Core expressions and so have interesting interactions with the invariants described above.

### 2.3 The advantage of a formalization gap

The goal of this work is to demonstrate the feasibility and benefits of *partial* verification. We wish to reason about some aspects of GHC while freely ignoring the rest. The reason for this attention focus is ultimately pragmatic: the implementation of GHC is big, complex, and not completely reconcilable with Gallina.

As one example, although the term language shown in Figure 2 is delightfully simple, the type and coercion languages are not. Representing these data structures in Gallina is challenging and reasoning about them would significantly increase our work. Fortunately, restricting our focus to the Expr data structure is achievable and justifiable; the operations that manipulate types and coercions are often independent of the term language operations, are implemented in separate modules in GHC, and maintain their own sets of invariants.

At a technical level, we implement this attention focus through the *edits* that guide and modify the hs-to-coq translation, as well as through the axiomatization of properties of part of the code base that we do not wish to reason about. We refer to this collection of edits and axiomatizations as the *formalization gap* that enables our work. In contrast to complete program verification, which seeks to minimize such differences, we embrace this feature as an enabling part of this work. We want to apply mechanical reasoning to novel code; a formalization gap makes that possible.

Section 6 discusses this formalization gap in more detail, along with our justifications for the consistency of the simplifications and axioms that we rely on. However, for context, we provide a preview of its main features here. For example, one goal of our case study is to focus our attention on Core terms while treating Core types and coercions abstractly.

|  | # Mods(hs) | LOC(hs) | LOC(v) |
|---|---|---|---|
| Handwritten Gallina | — | — | 429 |
| General purpose libraries | 12 | 1 270 | 2 279 |
| Compiler utilities | 9 | 7 478 | 4 962 |
| Core.v | 13 | 8 548 | 5 188 |
| Core passes | 15 | 5 268 | 5 431 |
| TOTAL | 49 | 22 564 | 18 289 |

**Table 1.** Translated part of GHC

Therefore, we model datatypes such as Coercion and Type_ using axioms.

```
Axiom Coercion : Type.
Axiom Type_    : Type.
```

Such axioms are a consistent addition to Coq's logic, as Type is inhabited. Furthermore, while we also axiomatize a few operations related to types and coercions, our proofs need no axioms about their properties.

Other notable simplifications that we make to the Core AST includer emoving the ability to represent type and coercion variables and the elimination of information from the AST related to passes we do not consider (i.e. unfoldings and rewrite rules). We also make assumptions about other parts of the implementation, including the properties of fresh name generation, free variable calculation, and finite sets of variables. As in the Coercion and Type_ axiomatization, we justify all axioms that we depend on for our reasoning. A complete list of these axioms appears in Appendix B.

The focused attention that these simplifications provide is essential. Although the operations that we ultimately reason about are a small part of GHC – the CoreSubst and Exitify modules are each under 500 lines of code – they have many dependencies. Table 1 presents the scale of our Gallina model. Notably, even after axiomatizing the the Type_ and Coercion data structures, our Gallina module defining the Core AST (called Core.v) still contains over 5000 lines of code. Furthermore, the definitions in this figure also rely on previously developed libraries from prior work: a translation of the GHC base library [49] and data structures for finite sets and maps from the Haskell containers library [11].

Finally, we note that the approach presented here is a *mechanized* formalization gap. By using hs-to-coq to construct our model of GHC, we gain three primary benefits over a model developed by hand.

**Mechanical assistance.** Developing a consistent mathematical model of a large and interwined code base is impractical. Although hs-to-coq must be guided via edits, it would be more work to define this model by hand.

```haskell
data Var
  = TyVar {    -- Type and kind variables
        varName    :: !Name,
        realUnique :: {-# UNPACK #-} !Int,
        varType    :: Kind }
  | TcTyVar {  -- Used only during type inference
        varName       :: !Name,
        realUnique    :: {-# UNPACK #-} !Int,
        varType       :: Kind,
        tc_tv_details :: TcTyVarDetails  }
  | Id {
        varName    :: !Name,
        realUnique :: {-# UNPACK #-} !Int,
        varType    :: Type,
        idScope    :: IdScope,
        id_details :: IdDetails,
        id_info    :: IdInfo }
```

```coq
Inductive ... (* part of a mutually inductive type *)
with Var : Type
  := | Mk_Id (varName    : Name.Name)
            (realUnique : BinNums.N)
            (varType    : Type_)
            (idScope    : IdScope)
            (id_details : IdDetails)
            (id_info    : IdInfo)  : Var
```

**Figure 3.** Haskell (top) and generated Gallina (bottom) definitions of Var.

- ***Richness.*** The model can be more detailed because it is developed by only eliminating parts that cause difficulties, as opposed to adding that which is *a priori* considered to be important.
- ***Provenance.*** The model is directly and observably connected to the actual implementation, and all simplifying assumptions are recorded in the edit files and can be inspected.

## 3 Formalizing the Core invariants

In this section we provide more detail about the two properties that we reason about in our work, namely the well-scopedness and join point invariants. However, before we do so, we must first discuss the representation of identifiers.

### 3.1 Representation of identifiers

Figure 3 presents the Haskell and Gallina representations of the Var type. Because we do not wish to reason about type information stored in the AST, we use edits to eliminate the TyVar and TcTyVar constructors as part of the simplification process. For a similar reason, we have also edited the IdDetails data type (whose definition is not shown) to eliminate the CoVarId constructor, which represents coercion variables.

GHC implements an efficient equality test between variables by comparing their *uniques*, which are the unboxed Int values stored in their realUnique fields. We use edits to modify the type of this field from Int to Coq's unbounded natural number type N, so that we need not reason about overflow. Despite the name realUnique, this integer is *not* guaranteed to be unique; multiple Vars may have the same unique but differ in their associated information (e.g., type, scope, and other details).

Therefore, we maintain the following invariant, called GoodVar, to enforce a dependency between the unique and two components of the associated information.

```coq
Definition GoodVar (v : Var) : Prop :=
  isLocalVar v = isLocalScope v /\
  varUnique v = nameUnique (varName v).
```

This invariant has two components. First, whether a variable is a local identifier or global identifier is determined in two ways: the bits of the unique itself provide this information in addition to the idScope component of the record. Our invariant checks if these two places are in sync. Although this relationship is not maintained by every pass of GHC (notably, the last Core-to-Core pass updates all scopes to global) it is maintained by our two passes of interest. Furthermore, by observing this relationship, we can simplify our reasoning, as we discuss in Section 6.5.

Second, the unique is actually stored in two places in the data structure: inside the name of the variable (varName) as well as in the realUnique field (accessed via the varUnique function). The second part of our invariant states that these two values should always be in sync.

### 3.2 Well-scoped expressions

We formalize the assertion that a Core expression is *well scoped* in Coq as a relation on a Core expression (i.e., a CoreExpr) and an in_scope set of type VarSet.[7]

```coq
WellScoped : CoreExpr -> VarSet -> Prop
```

Informally, an expression is well-scoped if every identifier in it is a GoodVar and if all local identifiers are contained within the current in_scope. The definition of WellScoped uses the predicate WellScopedVar, shown below, in order to enforce these restrictions on all variable occurrences in expressions.

```coq
Definition
  WellScopedVar (v : Var) (in_scope : VarSet) :=
    if isLocalVar v then
     match lookupVarSet in_scope v with
       | None => False
       | Some v' => almostEqual v v' /\ GoodVar v
     end
    else GoodVar v
```

---

[7]The complete Coq definition of this invariant appears in Appendix A.1.

In GHC, scopes are represented by VarSets. These VarSets are implemented by finite maps from uniques to Vars. Querying whether a variable is contained within this set is done by checking whether the unique of the variable is in the domain of the map. However, that query doesn't ensure that the *same* variable is stored in the set, only one with the same unique. Thus, we require a stronger property: not only should the variable stored in the set have the same unique, but it should also be the "same" variable in the sense of almostEqual, shown below. In that way, all of the occurrences of that variable in the expression will be forced to have the same meta-information.

The almostEqual propsition asserts that two variables differ only in their IdInfo and that all other components are identical. (In GHC, the the IdInfo component is used to store data that can be updated during an optimization phase; for example, strictness analysis can use it to record how a variable might be evaluated.)

```
Inductive almostEqual : Var -> Var -> Prop :=
 | AE_Id : forall n u ty ids idd id1 id2,
    almostEqual (Mk_Id n u ty ids idd id1)
                (Mk_Id n u ty ids idd id2).
```

In the case of binding (in Lam, Case and Let), the WellScoped predicate requires bound variables to be local, and extends the in_scope set with the new bound variables. There is no requirement for the bound variables to not already appear in the scope; the WellScoped predicate allows shadowing in expressions. In particular, a binder can be shadowed by another binder with the same unique, but perhaps different information (name, type, etc.). It is an explicitly documented requirement[8] that all passes must be able to handle input that has such shadowing.

### 3.3 Join points

Further invariants about the use of variables in GHC arise from *join points*, one of the most recent compilation innovations in GHC [40].

Join points are a way to express non-trivial local control flow (i.e., "jumps"). They are a more light-weight alternative to continuation-passing style. In existing paper formalizations [40], declaring join points and jumping to them are commonly their own syntactic categories. In GHC, the developers chose to represent them simply as normal let-bound function definitions and function calls, using a special marker on the function identifier (specifically, their IdDetail is a JoinId).

This leads to the following invariants surrounding the use of variables marked as join points, quoted directly from the GHC source code:[9]

---
[8]In Note [Shadowing] in the module CoreSyn.
[9]In Note [Invariants on join points] in the module CoreSyn.

1. *All occurrences must be tail calls. Each of these tail calls must pass the same number of arguments, counting both types and values; we call this the "join arity" (to distinguish from regular arity, which only counts values).*
2. *For join arity n, the right-hand side must begin with at least n lambdas. No ticks, no casts, just lambdas! C.f. CoreUtils.joinRhsArity.*
2a. *Moreover, this same constraint applies to any unfolding of the binder. Reason: if we want to push a continuation into the RHS we must push it into the unfolding as well.*
3. *If the binding is recursive, then all other bindings in the recursive group must also be join points.*
4. *The binding's type must not be polymorphic in its return type (as defined in Note [The polymorphism rule of join points]).*

We can formalize invariants 1, 2, and 3 in our setting. However, we cannot express either invariant 2a, because we edited away the unfoldings in IdInfo, or invariant 4, because we axiomatized all the Core type information.

In the course of doing our proofs, we found two further invariants that GHC maintains about join points:

1. The join arity must be non-negative.
2. A lambda-, case-, or pattern-bound variable cannot be a join point.

As before, we have a predicate, isJoinPointsValidProgram, that puts all this together and says when a complete Core program is valid. The Coq definition of this invariant appears in Appendix A.2.

## 4 Reasoning about Core

Our two main verification results are the theorems WellScoped_substExpr and exitifyProgram_WellScoped_JPV discussed in this section. These results themselves rely on many auxiliary lemmas that we do not describe here. Our Coq proof scripts (including the statements of the invariants above and all proofs) comprise over 13 000 lines of code. These results demonstrate that it is possible and informative to mechanically reason about the generated Gallina definitions that model the Core language.

### 4.1 Well-scoped substitution

The GHC implementation of substitution for core expressions is the function

```
substExpr : String -> Subst -> CoreExpr -> CoreExpr
```

that applies the given *substitution* (i.e., Subst) to an expression (i.e., CoreExpr), replacing multiple variables in parallel. The String argument provides documentation in the case of a scope violation; GHC dynamically checks the scope invariant during the operation of substitution. The Subst data structure is defined as follows:

```
Inductive Subst : Type
  := | Mk_Subst : InScopeSet -> IdSubstEnv
                  -> TvSubstEnv -> CvSubstEnv -> Subst.
```

This structure uses separate finite maps, called *substitution environments*, to record the individual substitutions for identifiers (IdSubstEnv), type variables (TvSubstEnv), and coercion variables (CvSubstEnv). In addition to these three environments, the substitution also maintains an InScopeSet: a set of variables that will be in scope *after* the substitution has been applied.

Because GHC uses a named representation of variables, the substExpr operation is careful to avoid capture by renaming bound variables using the following operation.

```
Definition substIdBndr : String -> Subst -> Subst ->
    Id -> (Subst * Id) :=
  fun _doc rec_subst '(Mk_Subst in_scope env tvs cvs
    as subst) old_id =>
    let old_ty := Id.idType old_id in
    let no_type_change :=
      orb (andb (isEmptyVarEnv tvs)
                (isEmptyVarEnv cvs)) true in
    let id1 := uniqAway in_scope old_id in
    let id2 := if no_type_change then id1 else
        Id.setIdType id1 (substTy subst old_ty) in
    let mb_new_info :=
        substIdInfo rec_subst id2 (idInfo id2) in
    let new_id :=
        Id.maybeModifyIdInfo mb_new_info id2 in
    let no_change := id1 == old_id in
    let new_env :=
      if no_change then delVarEnv env old_id else
        extendVarEnv env old_id (Mk_Var new_id) in
    pair (Mk_Subst (extendInScopeSet in_scope new_id)
            new_env tvs cvs) new_id.
```

This operation takes a documentation string (_doc), a recursive substitution (rec_subst), a substitution to apply (subst), and the original binding variable (old_id), and determines whether the original binding variable needs to be renamed. More specifically, it checks whether the identifier is already present in the in_scope set of the substitution (meaning that it could capture a free variable in the range of the substitution), and if so, renames the unique of the identifier using an operation called uniqAway. If the binding identifier was not renamed, then it is removed from the domain of the current substitution (cutting off further substitution for that variable). Otherwise, the renaming is added to the domain of the substitution. In either case, the binding identifier is added to the current set of in_scope identifiers.

Despite this identifier shuffling, we have shown that substitution is scope-preserving. Given a well-scoped substitution and a well-scoped expression, the result of applying that substitution is also well-scoped in the new scope indicated by the substitution.

**Theorem**
```
WellScoped_substExpr : forall e s expr_scope subst,
  WellScoped_Subst subst expr_scope ->
  WellScoped e expr_scope ->
  WellScoped (substExpr s subst e)
             (getSubstInScopeVars subst).
```

This theorem requires showing that the substitution subst satisfies the following invariant:[10]

> The in-scope set [of the substitution] contains at least those Ids and TyVars that will be in scope after applying the substitution to a term. Precisely, the in-scope set must be a superset of the free vars of the substitution range that might possibly clash with locally-bound variables in the thing being substituted in.

What this invariant means, in other words, is that the following two conditions must hold:

1. The in_scope set is a superset of the free variables of the expression minus the domain of the substitution.
2. The in_scope set is a superset of the free variables in the range of the substitution.

In our Gallina definition for the first condition we use the scope of the expression as an upper bound of its set of free variables, and interpret *superset* by a strong subset relation, written {<=}, that requires each variable of the first set to be almostEqual to some variable contained in the second. We ensure the second condition by requiring that expression in the range of the substitution to be well-scoped with respect to the in_scope set.

```
Definition WellScoped_Subst
    (s:Subst) (expr_scope:VarSet) := match s with
  | Mk_Subst in_scope subst_env _ _ =>
    minusDom expr_scope subst_env {<=}
      getInScopeVars in_scope
    /\ forall var,
      match lookupVarEnv subst_env var with
        | Some expr =>
          WellScoped expr (getInScopeVars in_scope)
        | None => True end end.
```

The most difficult part of proving the substitution lemma above is describing what happens when multiple binders (such as in a let expression) are potentially renamed by substIdBndr producing a new list of identifiers and a new substitution. In this case, we defined a SubstExtends property that describes the relationship that the original substitution

---
[10]Taken from a comment in the module CoreSubst.

s1 and list of binding variables vars1 have to the new substitution s2 and list of binding variables vars2.

This relation holds when

1. The lengths of the variable lists are the same;
2. There are no duplicates in vars2;
3. All of the vars in vars2 are GoodLocalVars;
4. The new variables vars2 do not appear in the in_scope set of s1;
5. The new in_scope set is strongly equal to the old in_scope set extended with the new variables;
6. The in_scope set, with the addition of the old variables, minus the domain of the new environment, is a subset of the new in_scope set; and
7. Anything in the new environment is either a renamed variable from the old environment or was already present.

These conditions are not present in the GHC source code, but are necessary to prove the well-scopedness property above. Together, they ensure that when multiple binders are renamed simultaneously, the invariants about binding are still preserved.

Verification did not reveal any bugs in GHC's definition of substitution, and we did not expect to see any in code this mature and well tested. However, this process has allowed us to precisely characterize the preconditions that guarantee a well-scoped result. Indeed, the specification of well-scoped substitutions was incorrect in the comments of GHC version 8.4.3 and was updated by the GHC developers to the description given above after correspondence with the authors.[11]

### 4.2 Exitification

When join points were added to GHC, they opened the door for new program transformations and simplifications [40]. One such opportunity is the ability to float a definition into a lambda.

Consider first the situation with regular functions:

```
let t = foo bar in
let f x = t (x*x) in
body mentioning f
```

It might be beneficial to inline t, replacing its occurrence in f with foo bar, as this can create new optimization opportunities in the body of f.

However, in general the compiler cannot do that in situations like this. As the code stands, t is evaluated at most once. If it was inlined into f, it would instead be evaluated as often as f is called. Thus, if t is expensive to compute, this could be an arbitrarily bad pessimization, and so GHC does not inline t.[12]

---

[11]https://gitlab.haskell.org/ghc/ghc/commit/
40d5b9e970149e85f0b5cbc1a795fa36f24a981a

[12]It *would* be safe to inline t, however, if the compiler knew that that f would be called at most once. And indeed there are multiple elaborate

The story is suddenly much simpler if f is not a general function, but actually a *non-recursive* join point j_f (we indicate join points with names starting with j_):

```
let t = foo bar in
let j_f x = t (x*x) in
body mentioning j_f
```

The join point invariants guarantee that all calls to j_f in the body are in tail-call positions. This implies that j_f is called at most once (more precisely: jumped to at most once), and the compiler may inline t at will.

This does not hold for recursive join points:

```
let t = foo bar in
let j_go 0 x y = t (x*x)
    j_go n x y = j_go (n-1) (x*x) (x+y)
in body mentioning j_go
```

Because j_go is *recursive*, its right-hand side will likely be evaluated multiple times, so inlining t would again risk repeated evaluation of t.

Or would it? Careful inspection reveals that in the case where t is evaluated, no further recursive calls to j_go occur. Or put differently: t is on the *exit path* of the loop represented by j_go, and inlining is safe after all.

The exitification optimization tries to find and recognize situations like these, and transforms the code so that it is obvious to the general purpose simplifier that t is used at most once. It does so by floating the expression on the exit path out of the recursive loop, into a non-recursive join point:

```
let t = foo bar in
let j_exit x = t (x*x) in
let j_go 0 x y = j_exit x
    j_go n x y = j_go (n-1) (x*x) (x+y)
in body mentioning j_go
```

Now we are again in the same situation we were with the non-recursive j_f before, and the simplifier will be able to inline t.

**Theorem proved** Any transformation that moves code from one scope to another needs to be careful about preserving the well-scopedness invariants, and exitification is no exception. It must delicately juggle names and scopes: the newly created exit join points must not shadow existing names, and they need a parameter for each free variable that is no longer in scope outside the recursive join point (x in this case, but not y). This made the proof that exitification preserves well-scopedness tricky.

Given that exitification only makes sense in the context of join points, and that it creates new join points, we were also naturally interested in knowing that the code that exitification produces adheres to the invariants about join points;

---

analyses in GHC that try to answer the question of whether f is called more than once [8, 47].

for example, that j_exit has the right number of lambdas and is invoked only from tail-call positions.

Furthermore, it turns out that the exitification code is actually not well-defined on Core terms that violate the join point invariants, so they also became a precondition for the well-scopedness proofs.

Therefore, we proved a combined theorem:

```
Theorem exitifyProgram_WellScoped_JPV: forall pgm,
  WellScopedProgram pgm ->
  isJoinPointsValidProgram pgm ->
  WellScopedProgram (exitifyProgram pgm) /\
    isJoinPointsValidProgram (exitifyProgram pgm).
```

***A bug is found*** While trying to show that exitification preserves well-scopedness, we found that it did not, at least in some obscure situations. Consider the following opportunity for floating out the expression foo x:

```
let j_go (x :: Bool) (x :: Int) =
    if x == 0 then foo x else j_go True (x - 1)
```

Note that the second parameter of j_go shadows the first. The previous version of exitification abstracted the exit expression over all locally bound variables:

```
let j_exit (x :: Bool) (x :: Int) = foo x in
let j_go (x :: Bool) (x :: Int) =
    if x == 0 then j_exit x x else j_go True (x - 1)
```

But this is now wrong: The first argument to j_exit should be of type Bool, but a value of type Int is used instead!

Under normal circumstances, the Core passed to the exitification pass does not exhibit such shadowing. Nevertheless, it is a bug that was worth fixing.[13] In particular, users of GHC may insert custom Core-to-Core passes at any point in the pipeline via compiler plugins, and such plugins are becoming more popular [4, 9, 20, 22, 24, 36]. GHC must be able to handle any possible shadowing in Core in case such passes introduce them.

Because of this bug, the version of the Exitify module that we currently verify is from GHC 8.6.1, not 8.4.3.

## 5 Extending the `hs-to-coq` edit language

As part of developing our translation of GHC, we have added new functionality to hs-to-coq in support of this mode of use. In this section, we describe these features in more detail.

***Existing capabilities of `hs-to-coq` edits*** To provide context to our new additions, we first summarize the existing capabilities of hs-to-coq edits and their typical uses in translation. Prior work on translating the containers library demonstrated the use of edits for medium-scale programming [11]. At this scale, it was already necessary for users of the tool to define a number of edits in order to translate this library. We can categorize these edits as *reconciliation*, *selection*, or *simplification* edits.

---

[13]GHC issue #15110, fix first released with GHC 8.6.1.

Typical reconciliation edits, which align the semantics of Haskell and Gallina, included:

- Removing Haskell features that have no counterpart in Gallina, including reallyUnsafePtrEquality# and seq, using **rewrite** edits.
- Managing recursive functions that were not structurally recursive by either providing the termination proof or by deferring those proofs altogether, using **termination** edits.
- Replacing operations that throw exceptions, such as error, with default values, via **rewrite** edits. These default values are propagated via the Default type class and are used opaquely.

Typical selection edits, which focus on specific parts of the code base, included:

- Skipping parts of the code that are irrelevant for verification or are difficult to translate (such as code related to serialization and deserialization), using **skip** edits.

Typical simplification edits, which make the code easier to reason about, included:

- Modifying the representation of integers to avoid reasoning about overflow, using **rename type** edits.
- Substituting operations that are difficult to reason about (e.g., bit twiddling functions) with simpler definitions, using **redefine** edits.
- Replacing small code fragments with alternative expressions, using **rewrite** edits.

Our translation of GHC requires all of these previously-extant forms of edits. However, because we found that this functionality was not enough, we also extended hs-to-coq with new edit forms, described below. In the next section, we provide further details about the specific use we make of them when translating GHC.

***Simplification: Constructor skipping*** The Core AST, though simpler than source Haskell, carries around a great deal of information that is not germane to our verification goals. Some of this information is in the form of metadata; some of it is in the form of type and coercion variables that we do not analyze. Regardless, we would like to avoid dealing with these concerns. However, the problematic cases are often *sub*cases of other data types – for example, the Expr data type for Core contains a case Tick strictly for profiling information, as we see in Figure 2. Thus, we added support for a new **skip constructor** edit that eliminates an entire case from data types and then propagates this information to delete any equation of a function definition or arm of a case statement that matches against this constructor.

For example, the edit **skip constructor** Core.Tick removes the Tick constructor from of Expr. More dramatically,

we use this edit to modify the representation of variables, which we discuss further in Section 6.1.

In this way, the embedding itself provides assurance that our code of interest is independent of particular features of GHC, without any proofs. In particular, if the targeted code needed to use the skipped constructor in some fundamental way (e.g., if it were used to construct a value in an operation that could not be skipped), then the output of hs-to-coq would not be accepted by Coq. We can only skip constructors that we can isolate from the rest of the development.

We do need to be careful, however – heavy use of **skip constructor** can lead to wildcard cases that no longer match anything, as all the would-be "extra" constructors have been skipped. Because Coq does not permit redundant pattern matches, we also add an edit to manually delete such cases. The edit **skip equation** f pat1 pat2 ... removes the equation matching pat1 pat2 ... from the definition of the function f.

***Selection: Axiomatization***   The ability to axiomatize definitions and modules was added to hs-to-coq directly for this project. There are many definitions in GHC that we want to reason about abstractly. These include code that is used within the other functions that we want to verify, but not on a code path that is exercised by our proof (for example, to manipulate the metadata stored with expressions). We thus provide an **axiomatize definition** edit, which replaces a definition in Haskell with a Coq **Axiom** at the same type.

We offer multiple ways to interact with axiomatization. For example, while we are focused on Core, its dependencies transitively reach into much of GHC, and we don't want to deal with all of them. While some modules we can skip (via **skip module**), this isn't always viable. For example, the FastStringEnv module declares a type for maps keyed by GHC's FastString type. These are used, for example, when manipulating metadata for data type constructors, but this is not an operation we need to concern ourselves with to verify properties of variables. We can thus **axiomatize module** FastStringEnv, which leaves the type definitions intact and automatically axiomatizes every definition in the module as per **axiomatize definition**.

We also use axioms to replace *type* definitions. As type definitions do not have kind annotations in Haskell, we cannot automatically generate axioms; we instead use the **redefine** or **rename** edits to replace one definition with another. For example, **redefine** Axiom DynFlags.DynFlags : Type replaces a record of configuration options with an opaque axiom.

While being able to axiomatize Haskell definitions is important, it does have the potential to introduce inconsistency – if we axiomatized the Haskell definition undefined :: a, we would be able to prove any theorem we wanted. As a result we need to examine the functions we axiomatize; however, in GHC, most functions are not fully polymorphic and return inhabited types. Appendix B lists the relevant axioms for our development and the justification of their consistency with Coq.

We discuss the use of axiomatization further in Section 6.0.1, with a focus on its specific importance for extracting a slice of GHC. However, even though we automatically axiomatize many definitions in our development, we almost never assume axioms about their properties. We discuss why this is the case in Section 6.

***Reconciliation: Type modification***   Prior work either avoided partial operations altogether [49] or attempted to isolate them behind total interfaces [11]. That isn't possible with GHC – many more operations may fail, for many reasons.

One source of partiality comes from the use of GHC's operation for signaling a run-time error (i.e., a compiler bug) – the function Panic.panic. We cannot translate this function, since it actually throws an exception (using unsafeDupablePerformIO, no less). Instead, we axiomatize this operation as follows:

```
Axiom panic : forall {a} `{GHC.Err.Default a}, GHC.
    Base.String -> a.
```

This constraint therefore enforces that panic can only be called with a return type that is known to be inhabited, ensuring that it does not introduce unsoundness. Although panic does not terminate the entire Coq program as it does in Haskell, arriving at it in a proof terminates our ability to reason about the code. Therefore, proving properties about code that uses panic also increases our confidence that it will not be triggered on that code path. For example, GHC uses Haskell's ability to define record selectors for single constructors of data types with multiple branches; these selectors are necessarily partial.

While the Default class is not new, we have made more significant use of it here than in prior work. As a result, we sometimes need to add Default constraints to types where they weren't already present. To guide this translation, we introduce a new **set type** edit, which allows us to change the type of a definition to a new type of our choosing. It is always safe to use this edit, as Coq's typechecker will prevent us from assigning an inconsistent type to a definition.

***Reconciliation: Mutually recursive modules***   GHC, nearly uniquely among Haskell programs, makes significant use of *recursive* modules; most of the modules that define Core are part of a single mutually-recursive cycle. This is not a feature supported by Coq, so as part of the translation, we have had to introduce edits that combine multiple source modules (both translated and axiomatized) into a single target. We use this facility to create the module Core, which contains the definition of the abstract syntax of the Core language.

***Reconciliation: Mutual recursion edits***  The Core AST that we saw in Figure 2 is defined via the mutually inductive types `Expr` and `Bind`. Consequently, most operations that work with either of these data structures are mutually recursive. For example, consider the `exprSize` operation, shown in the left hand side of Figure 4; it is mutually recursive with the analogous function `bindSize`. These functions compute the size of an expression and a binder, respectively. Coq can natively show the termination of mutually defined fixed points as long as they are each recursive on one of the mutually recursive types and make recursive calls to each other on strict subterms of their arguments. One important pattern this naive treatment of termination prohibits is "preprocessing" – recursion that goes through an extra function which simply forwards to one of the recursive functions. This is something we see in as simple a definition as that of `exprSize` and `bindSize`. Along the way, the function `altSize` is called, which simply unpacks a tuple and recurses into `exprSize` and `bindSize`. From Coq's perspective, that means *all three* of these functions are mutually recursive, and one of them has as its only argument a tuple. And there's a fourth function, `pairSize`, which has the same problem.

Since a tuple isn't a mutually recursive inductive data type, for Coq to accept the definitions of `exprSize` et al., we must *inline* the definitions of `pairSize` and `altSize` into the mutually recursive definitions that use them. To tell `hs-to-coq` to do this, we use the **inline mutual** edit:

**inline mutual** CoreStats.pairSize
**inline mutual** CoreStats.altSize

This results in the Coq definition on the right in Figure 4, as well as new free-standing *non*-recursive definitions of `pairSize` and `altSize` which are the same as their (local) `let`-bound definitions.

***Reconciliation: Type inference***  Haskell's ability to perform type inference is significantly stronger than Coq's, particularly for program fragments that remain (as many do) within the bounds of Hindley-Milner type inference. For the most part, Coq's type inferencer is powerful enough, when combined with the presence of type annotations on top-level bindings, to infer all the types we need. There are, however, occasional exceptions. One subtle case is that Coq cannot infer a polymorphic type without explicit binders, as it cannot insert binders for type variables automatically.

In order to work around this, we augmented `hs-to-coq` in two ways. One is the above **set type** edit, which allowed us to monomorphize local functions that could have been polymorphic but were only ever used at one type. The other is that we taught `hs-to-coq` to annotate the binders of a fixpoint with their types: to go from `let f : A -> B := fix f x := ... in ...` to `let f : A -> B := fix f (x : A) : B := ... in ...`. Without this transformation, Coq's type inferencer would sometimes fail to infer the type of a fixpoint, as the type information was too far away.

## 6 A mechanized formalization gap

There are two forms of formalization gap in our work: the gap introduced by the translation itself, that lies in the difference in semantics between the Haskell and Gallina versions of GHC; and the gap that derives from simplifications introduced by edits and the introduction of axioms in our proofs.

Pragmatically, we cannot work without either. Because Haskell and Coq do not have the same semantics, and because GHC takes advantage of Haskell idioms that are difficult to map into Coq, we will always have some sort of reconciliation gap. The second form of gap is also important in terms of pragmatic proof development. We want to reason about the most interesting parts of the code base first and defer the verification of other parts, perhaps indefinitely.

In this section, we describe this second form of gap in more detail, including the application of edits and the axioms that we assume as part of our Core language specification and proofs. Developing this translation was a significant part of this project, so we view it as a separate contribution of our work. The translation itself is guided by 1411 lines of edits and 343 lines of inserted Gallina code.

When constructing the edits that guide this translation, we follow the general design principle of "make illegal states unrepresentable"[14]. In other words, we set up our edits so that situations that we do not want to reason about are eliminated from the translation. As much as possible, we try to use (nondependent) types to capture invariants about the data structures that we reason about.

### 6.0.1 Removing dependencies through axiomatization

Although the Core data type and operations that we target are a small part of GHC, they have many dependencies on modules throughout the compiler. These dependencies are an issue because the translation of Haskell code to Gallina is not fully automatic. This is especially true for GHC; we have found that almost every module we have translated requires custom edits. Because there is a cost to developing the edits necessary for the translation, we would like to do as little of it as necessary. We don't want to waste time figuring out how to translate Haskell code that we are uninterested in reasoning about.

Fortunately, we are not starting from scratch. Our embedding of GHC's source code relies on previously developed libraries from prior work, including a translation of the GHC `base` library [49] and data structures for finite sets and maps from the `containers` library [11].

---

[14]Yaron Minsky, "Effective ML", https://blog.janestreet.com/effective-ml-video/

```
exprSize :: CoreExpr -> Int                              Definition exprSize : CoreExpr -> nat :=
-- ˆ A measure of the size of the expressions,             fix exprSize (arg_0__ : CoreExpr) : nat :=
-- strictly greater than 0                                   let altSize (arg_0__ : CoreAlt) : nat :=
exprSize (Var _)        = 1                                        let 'pair (pair _ bs) e := arg_0__ in
exprSize (Lit _)        = 1                                          bndrsSize bs + exprSize e
exprSize (App f a)      = exprSize f + exprSize a        in match arg_0__ with
exprSize (Lam b e)      = bndrSize b + exprSize e          | Mk_Var _ => 1
exprSize (Let b e)      = bindSize b + exprSize e          | Lit _ => 1
exprSize (Case e b _ as) =                                 | App f a => exprSize f + exprSize a
  exprSize e + bndrSize b + 1 + sum (map altSize as)       | Lam b e => bndrSize b + exprSize e
exprSize (Cast e _)     = 1 + exprSize e                   | Let b e => bindSize b + exprSize e
exprSize (Tick n e)     = tickSize n + exprSize e          | Case e b _ as_ =>
exprSize (Type _)       = 1                                   ((exprSize e + bndrSize b) + 1) + sum (map altSize as_)
exprSize (Coercion _)   = 1                                | Cast e _ => 1 + exprSize e
                                                           | Mk_Type _ => 1
bindSize :: CoreBind -> Int                                | Mk_Coercion _ => 1
bindSize (NonRec b e) = bndrSize b + exprSize e            end
bindSize (Rec prs)    = sum (map pairSize prs)           with bindSize (arg_0__ : CoreBind) : nat :=
                                                           let pairSize (arg_0__ : (Var * CoreExpr)%type) : nat :=
pairSize :: (Var, CoreExpr) -> Int                               let 'pair b e := arg_0__ in
pairSize (b,e) = bndrSize b + exprSize e                           bndrSize b + exprSize e
                                                         in match arg_0__ with
altSize :: CoreAlt -> Int                                  | NonRec b e => bndrSize b + exprSize e
altSize (_,bs,e) = bndrsSize bs + exprSize e               | Rec prs => sum (map pairSize prs)
                                                           end
                                                         for exprSize.
```

**Figure 4.** Mutual recursion in Haskell (left) and Gallina (right)

**Handwritten Gallina:** AxiomatizedTypes, FastString, NestedRecursionHelpers, IntMap
**General purpose libraries:** Bag, EnumSet (ax), BooleanFormula, UniqFM, UniqSet, OrdList, FiniteMap, ListSetOps, Maybes, MonadUtils (ax), Pair, State
**Compiler utilities:** Util, SrcLoc, Unique, UniqSupply, BasicTypes, DynFlags (ax), Panic (ax), OccName, Module,
IdInfo, Class (ax), TyCon (ax), DataCon (ax), PatSyn (ax), Var, VarSet, VarEnv, CoreSyn, Demand (ax), Type (ax), TyCoRep (ax), Coercion (ax)
**Core.Core operations and passes:** FastStringEnv (ax), Constants (ax), Id, PrelNames, CoreUtils, Name, NameEnv, NameSet, FV, Literal (ax), FieldLabel (ax), ConLike (ax), CoreFVs, CoreSubst, Exitify

Axiomatized modules marked with (ax).
**Table 2.** Translated GHC modules

These libraries are already a significant starting point. For example, the base library contains 40 Coq modules and 8834 non-blank, non-comment, lines of Gallina definitions. Similarly, the containers library contains 13 Coq modules and 7492 lines of Gallina definitions. Each of these examples are as faithful translations of the Haskell versions as possible. While it was important that the edit files could modify the translation in support of verification, the goal for these libraries was for the semantics of the Coq output to match the Haskell implementation. In particular, extracting the Coq version of the containers library back to Haskell allows it to pass the original test suite from the Haskell library, demonstrating that the Coq version has the same semantics as the original.

Even though we do not need to worry about dependencies on the base libraries, we do have the issue of dependencies within GHC itself. Because GHC is a large software system, its dependencies are both complex and deep – we don't want to be forced to start at the leaves of the hierarchy in our translation, as the code we are interested in is somewhere in the middle of the dependency graph. For example, the CoreSyn module, which contains the AST shown in Figure 2, imports 23 different GHC-internal modules. Many of these we would also like to reason about in our formalization (e.g., VarEnv, VarSet, etc.) because they directly relate to the representation of Core terms and variables. However, this module also refers to functions and types defined in less pertinent modules including CostCentre (profiling information) and Outputable (formatting error messages). Some of these imported modules we only want to reason about abstractly. For example, we don't care how the DynFlags module represents compiler options, but we do need to know what the options are and how they may interact with compilation passes.

To avoid this extra complexity we make heavy use of **skip** and **axiomatize** edits in our translation. For example, of the 42 modules imported by the DynFlags module, only ten remain after axiomatization: five from the base libraries, one from the containers library and only four modules from GHC. As a result of these edits, our translation of the Core language draws from a total of 49 modules of GHC, which themselves contain over 22 000 lines of Haskell code. These modules are summarized in Table 1.

Another issue with defining the translation is GHC's use of recursive modules, as mentioned in Section 5. Of our 49 identified modules of interest, half of them belong to a single mutually recursive module group, whose structure is shown in Figure 5. While our edits allow us to merge these disparate Haskell modules into a single Gallina module, which we call Core, we don't want to deal with the entire module graph. Therefore we carefully axiomatize and skip definitions to successfully break the cycles (axiomatizing types and coercions is a particular help) and reduce the size of this aggregate module.

### 6.1 Eliminating Core type and coercion variables

As mentioned in Section 2.2 we use **skip constructor** edits to eliminate the representation of type and coercion variables from GHC. We also axiomatize the definition of the Type_ and Coercion data structures.

With this modification, we would be justified in adding axioms to our proof development that assert that there are no free variables to be found in types and coercions. (As we axiomatize the functions that calculate free type and coercion variables, tyCoFVsOfType and tyCoFVsOfCo, we cannot prove or disprove these properties in Coq.)

```
Axiom tyCoFvsOfType_empty : forall ty,
  TyCoRep.tyCoFVsOfType ty = FV.emptyFV.
Axiom tyCoFvsOfCo_empty   : forall co,
  TyCoRep.tyCoFVsOfCo co   = FV.emptyFV.
```

This approach would work, but our proofs are simpler if we replace these axioms with edits. We therefore use **rewrite** edits to instruct hs-to-coq to immediately replace these function calls with an empty set of free variables as it translates the Haskell code:

```
rewrite forall ty,
  TyCoRep.tyCoFVsOfType ty = FV.emptyFV
rewrite forall co,
  TyCoRep.tyCoFVsOfType co = FV.emptyFV
```

This translation strategy has its advantages. In terms of proof, it is more pragmatic as the property is applied automatically. We can save our time for the details that matter. Furthermore, edits are generally safer than axioms, as – although they can be wrong – they cannot introduce unsoundness into Coq's logic.[15] Even so, there is a cost to using a **rewrite** edit instead of an axiom: the rewrite leaves no trace in the generated code. If we had used the axioms instead, we would be able to see where this sort of reasoning is required, potentially leading to more robust proofs. Furthermore, for each theorem, Coq can list the axioms that it depends on. We have no way to track such dependencies on rewrites.

---

[15]Rewrites can only turn Gallina terms into other Gallina terms. As long as the result compiles, it is logically consistent.

### 6.2 Eliminating coinduction

We model the Core data type (Figure 2) as an *inductive* data type. However, because Haskell is a nonstrict language, this interpretation is not quite accurate – it is a coinductive structures in Haskell. However an inductive interpretation of Core seems reasonable – Haskell programs are finite, after all, so they should be representable using a finite AST. And it is almost true for Expr.

Surprisingly, GHC treats Expr as a mixed inductive/coinductive data type. At first, the parsed and type-checked AST is finite. But, during compilation, GHC augments identifiers with additional information about *unfoldings* (the identifier's right-hand side that may replace its occurrence) and *rules* (possible context-dependent rewrites that the optimizer can apply [45]). The programmer can specify these optimizations through pragmas, or the compiler can create them on its own (for example, small functions tend to be inlined automatically, and GHC uses rules to specialize class methods).

This information is stored and manipulated coinductively in GHC – for instance, if a variable appears in a recursive binding, then its unfolding is an expression that may include a reference to that same variable. At the same time, the use of coinduction is limited to this sort of metadata. For example, the GHC developers expect the exprSize function (which ignores information attached to identifiers) to terminate Figure 4).

While hs-to-coq could be directed to represent the Expr data type as a coinductive type, this would be disastrous. Coinductive data structures can only be eliminated to produce other coinductive data structures. We would not be able to show that many perfectly reasonable operations terminate, such as exprSize, and we would not be able to use induction in our proofs. Therefore, to allow a fully inductive interpretation of Core, we use edits to drop information about unfolding and rewrite rules from the data type. For rewrite rules, we replace the RuleInfo data type with a trivial one:

```
redefine Inductive Core.RuleInfo : Type :=
  Core.EmptyRuleInfo.
```

We also redefine operations that work with RuleInfo, reflecting that our translated version of GHC is not allowed to include any information about rewriting rules.

```
redefine Definition Core.isEmptyRuleInfo
  : Core.RuleInfo -> bool
  := fun x => true.
```

For Unfolding, we use the **skip constructor** edit to eliminate every constructor of the data type except the no-argument NoUnfolding constructor.

Furthermore, Haskell functions that work with the Core AST also need to be edited when they use knot-tying definitions to process this coinductive data. For example, in substitution, the result that is produced when traversing a list of recursive binders is itself defined via corecursion by passing a recursive occurrence to substIdBndr.

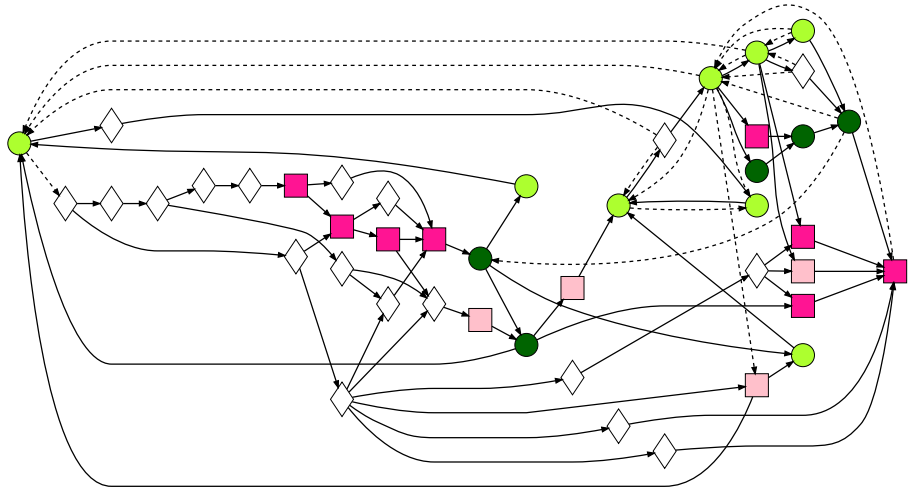

**Figure 5.** Mutually recursive modules related to the Core intermediate language in GHC. Each node is a GHC module. Solid lines are direct imports, dotted lines are "source" imports (how GHC Haskell marks recursive module imports). Green circles are modules that become part of the Gallina Core module, darker for translated modules and lighter for axiomatized ones. Pink squares are modules that become ordinary standalone modules, darker for translated modules and lighter for axiomatized ones. White diamonds are skipped modules.

```
-- Substitute in a mutually recursive group of 'Id's
substRecBndrs :: Subst -> [Id] -> (Subst, [Id])
substRecBndrs subst bndrs = (new_subst, new_bndrs)
  where (new_subst, new_bndrs) =
          mapAccumL (substIdBndr (text "rec-bndr")
                    new_subst) subst bndrs
```

However, the recursive substitution argument passed to substIdBndr is only used to update the metadata – specifically rules and unfoldings, which we have made trivial. Thus our translation of the substIdBndr function will never need it. Instead, we can pass any well-typed term in its place, and so we use the translation of the Haskell error function instead.

```
in CoreSubst.substRecBndrs
  rewrite forall x,
    CoreSubst.substIdBndr x new_subst =
      CoreSubst.substIdBndr x
        (GHC.Err.error Panic.someSDoc)
```

This edit allows us to produce the following definition for substRecBndrs, which is neither recursive nor corecursive:

```
Definition substRecBndrs
  : Subst -> list Core.Id -> (Subst * list Core.Id)%
    type :=
```

```
fun subst bndrs =>
  let 'pair new_subst new_bndrs :=
    mapAccumL (substIdBndr (Datatypes.id (GHC.Base.
  hs_string__ "rec-bndr"))
                           (GHC.Err.error Panic.
  someSDoc))
             subst bndrs in
  pair new_subst new_bndrs.
```

Why is this edit justified? First, because as previously mentioned, we have used our edits to remove all expressions that occur in the metadata. Second, because if we ever actually need to use the recursive substitution argument in our proofs, we will just find error instead.

### 6.3 Replacing data structures

Parts of the GHC code base depend on data structures from Haskell's containers library. For example, the types VarSet, DVarSet, VarEnv, and DVarEnv for variable sets and environments are implemented via the containers data structure Data.IntMap. As a result, reasoning about GHC requires understanding the properties of this data structures. But, as our interest is in GHC, we would rather not spend time on Data.IntMap. Although some parts of the containers library have been verified [11], Data.IntMap has not.

One approach is to axiomatize the properties of the finite map structure that we need for the proof. But, how would we know that this axiomatization is consistent? It turns out that the related containers data structure Data.Map has been proven correct. While this is not the finite map library used by GHC, it does have the same interface as Data.IntMap. Therefore, we can increase the confidence in our work by using edits to replace all uses of Data.IntMap with Data.Map, and justifying the required finite map properties with the theorems of Data.Map.

Furthermore, we use data structure replacement via edits to model DVarSet and DVarEnv as well. In this case, the two modules have almost the same interface as VarSet and VarEnv; the difference between them is the order that elements are iterated over during folds, and the fold operations are never used in our modules of interest. Therefore, our edits reimplement DVarSet as VarSet and DVarEnv as VarEnv, further allowing us to avoid uninteresting verification.

### 6.4 Valid VarSets

As described above, VarSets are implemented by finite maps in GHC. A set of variables is a finite map from a variable's unique (i.e., an integer) to the variable itself. The source code of VarSet records an invariant about this representation: *if a VarSet maps a unique key to a Var, then that key must be the same as the one stored in the Var.*

```
lookupVarSet :: VarSet -> Var -> Maybe Var
  -- Returns the set element, which may be
  -- (==) to the argument, but not the same as
```

We need to use this property in our proofs, so we add the following axiom that states that all VarSets satisfy the required property.

**Definition** ValidVarSet (vs : VarSet) : Prop :=
　forall v1 v2,
　　lookupVarSet vs v1 = Some v2 -> (v1 == v2).
**Axiom** ValidVarSet_Axiom : forall vs, ValidVarSet vs.

This axiom does not hold of all VarSets, but we are confident that it holds of all VarSets used in GHC. In other words, we have proven that this property is an invariant of the VarSet implementation, and GHC defines VarSets using an abstract type so that this invariant can be preserved.

We could avoid this axiom (in future work) by defining VarSets using a sigma type of the finite map paired with the invariant above. In that way, clients of the type would be able to access the representation invariant. However, while we have constructed such a definition of VarSet by hand, an automatic translation is beyond the scope of hs-to-coq.

### 6.5 Free variables and exitify

GHC knows two ways to calculate the set of free variables of a Core expression:

```
exprFreeVars :: CoreExpr -> VarSet
freeVars     :: CoreExpr -> CoreExprWithFVs
```

The former simply calculates the set of free variables, while the latter copies the Core expression and annotates all its subexpressions with their free variables. From a CoreExprWithFVs we can get this annotation with freeVarsOf. The function deAnnotate strips the annotation.

In the proofs about the exitify transformation (Section 4.2), both ways come up and we need to relate them. We could conveniently assume an axiom that states that the produced sets are equal, but this property is not quite true. The internal structure of the sets could differ. We could also assert, with a better conscience, that the two sets *denote* the same sets, but that would entail some rather tedious proofs that the exitify code respects this equivalence.

Therefore, instead of using an axiom, we use a **rewrite** edit to express that, in the context of the exitify pass, the directly calculated free variables set can be used instead of the annotation:

```
in Exitify.exitifyRec rewrite forall ann_e,
   freeVarsOf ann_e = exprFreeVars (deAnnotate ann_e)
```

As an axiom, this equation would be unsound, as the annotation in an CoreExprWithFVs could in general be anything. But as a rewrite edit, it is justifiable: We know exactly in which context the rewrite is applied, and we only assume that this equation holds in this particular context.

Another property that exitify requires of exprFreeVars is that, if an expression is well scoped, its free variables are a subset of that scope. In our proofs, we state and prove the following property using the subVarSet operation (translated from GHC) that compares two sets in terms of their uniques.

**Lemma** WellScoped_subset:
　forall e vs, WellScoped e vs ->
　　subVarSet (exprFreeVars e) vs = true.

However, to prove this property, we make a very subtle simplification in our translation. The exprFreeVars function works by calculating *all* of the free variables of an expression but then filtering that set so only the local variables remain, using isLocalId function. In GHC, this function looks at the idScope in the Var to determine whether a variable is a local var. However, in our Gallina version, we redefine this function so that it makes the decision based on the unique instead. As long as these two components stay in sync, i.e. as long as our GoodVar predicate holds, this change makes no difference to the behavior of the function.

### 6.6 Uniques and uniqAway

One operation that we axiomatize is the uniqAway operation used to produce fresh variable names.

**Axiom** uniqAway : InScopeSet -> Var -> Var.

In GHC, this operation tries to find a fresh variable for a given InScopeSet by repeatedly guessing. If it cannot find one after one thousand guesses, the operation gives up and panics. The correctness of substitution and exitify depends on the uniqAway operation always successfully producing fresh variables. However, there is no guarantee that one thousand guesses will be enough.

Because we don't want to introduce probabilistic reasoning, we therefore ignore uniqAway's (translatable) implementation and axiomatize it along with the properties that we require. In particular, we rely on axioms that state that uniqAway...

- returns a variable that is not in scope...

  ```
  Axiom uniqAway_lookupVarSet_fresh : forall v
      in_scope_set,
    lookupVarSet (getInScopeVars in_scope_set) (
    uniqAway in_scope_set v) = None.
  ```

- preserves the invariants of good variables...

  ```
  Axiom nameUnique_varName_uniqAway:
    forall vss v,
    nameUnique (varName v) = varUnique v ->
    nameUnique (varName (uniqAway vss v)) =
      varUnique (uniqAway vss v).
  Axiom isLocalId_uniqAway : forall v iss,
    forall iss v,
    isLocalId (uniqAway iss v) = isLocalId v.
  ```

- and any information associated with the variables...

  ```
  Axiom id_details_uniqAway: forall iss v,
    id_details v = id_details (uniqAway iss v).
  Axiom idScope_uniqAway:    forall iss v, idScope
      v   = idScope (uniqAway iss v).
  ```

- and that it doesn't modify variables that don't need to be freshened.

  ```
  Axiom uniqAway_eq_same : forall v in_scope_set,
    (uniqAway in_scope_set v == v) -> (uniqAway
    in_scope_set v = v).
  ```

## 7 Related work

**Liquid Haskell**  Liquid Haskell [53] also shares the goal of verifying real-world Haskell code. Breitner et al. [11] extensively discusses the relation between Liquid Haskell and prior uses of hs-to-coq as well as comparisons to other tools and methodologies for verifying Haskell programs. Our work extends the scale of verification relative to these systems, demonstrating that mechanical reasoning through shallow embeddings is possible for code extracted from a system with more than a hundred thousand lines of code.

**Refinement**  Refinement relations, along with forward and backward simulation relations that generalize them, are commonly used to describe a correspondence between two programs doing the same "important" computations [1, 37, 42]. For example, seL4 uses refinements to connect kernel code written in C with compiled binary code, and to extend verification done on the former to the latter [48]. CompCert uses backward simulation to express its semantic preservation theorem [35].

Refinement relations are also commonly used to establish a correspondence between a concrete implementation and a specification for which formal analysis can be more easily performed. For example, seL4 defines an abstract model of its operating system and bases its verification on this. Refinements then show that all the Hoare logic properties of the abstract model also hold for the kernel source code [30]. CertiKOS also uses simulations to construct certified abstract layers to facilitate modular reasoning [25]. The usage of refinement relations is also popular for reasoning about concurrent or distributed systems [26–28, 31, 33] and it has been shown that observational refinement is equivalent to linearizability, a popular correctness condition for concurrent systems [23].

Refinement has also been used from the other direction in program development: the process starts from a high-level specification and then applies several refinement steps to derive a concrete implementation [14, 16, 17, 54]. A recent success story of this approach is a high performance implementation of an elliptic-curve library that has been deployed to Chrome, Android, and CloudFlare [21].

There is a superficial similarity between refinement and this work. In both cases there is a connection between two versions of a system, where one form is more suitable for mechanical verification. However, whereas refinement proves the equivalence between the two systems so that the proofs directly carry over, our work merely establishes a relationship (the edit-based translation) and does not attempt to reason about it. Furthermore, with the exception of synthesis (e.g., Fiat [16]), with refinement the two systems are constructed by hand, while we use an automated transformation.

Our overall process does not eliminate the possibility of the development of more precise characterizations of the compiler implementation in future work. Indeed, the recorded simplifications of our edit files could be used to express the refinement relation with a future, more detailed model.

**Prior work on verifying compilers**  Compilers play critical roles in any software development, and therefore it is important to ensure that they are correct. Despite decades of effort invested in testing compilers, we are still far from the ideal. Past work on mechanical verification for GHC has focused on verifying descriptions of parts of the compiler developed by hand [6, 7]. These models simplify and elide many details that appear in the implementation. How do we know what simplifications have been made? How do we know how well these models correspond to the algorithms that GHC implements?

Mechanical verification has been an important component in the development of new compilers, starting with the ground-breaking CompCert, a formally verified optimizing C compiler implemented in Coq [34]. This approach has been shown to be effective: the study by Yang et al. [55] failed to find any wrong-code errors in CompCert despite devoting significant resource to the task. Other inspiring endeavors in this direction include Vellvm, which formalizes the operational semantics of LLVM IR and its SSA-based optimizations in Coq [56, 57]; CakeML, a formally verified ML compiler implemented in HOL4 [51]; and CertiCoq, a verified optimizing compiler for Coq [2].

Our work differs from these impressive results because we target the actual source code of GHC, a mainstream, industrial-strength compiler that was written without verification in mind. The advantage is that the verification is separate from the development: the compiler developers don't have to know formal reasoning techniques, and verification does not interact with their workflow. In contrast, the systems above are based on new implementations that were developed in conjunction with their verification. While we agree that this approach is more likely to lead to completely verified software systems, it does not enable mechanically-assisted reasoning for existing code.

***Prior work on variable binding***  Our proofs in GHC are mainly about the use of variables in the abstract syntax of lambda-calculus based intermediate languages. Many mechanized developments of lambda terms select variable representations, such as de Bruijn indices [46], locally nameless representations [41], higher-order abstract syntax [13, 18], or nominal logic [52], that simplify this reasoning. Of the solutions presented to the POPLmark challenge [3] only one (by Aaron Stump) used a fully named representation of binding like GHC does.

The representation of variables in verified compilers often varies. CakeML uses strings for variables at the source level but uses closures instead of substitution to describe their semantics. The compiler then translates this representation to an intermediate language that uses de Bruijn indices [32, 51]. CertiCoq works in the opposite way, relying on a de Bruijn representation for the AST of Gallina, and then translating to a named representation for the CPS conversion [2].

## 8 Future work and conclusions

Although our goal has been to demonstrate that partial verification is possible and useful for a system like GHC, there is of course more that could be proved using the model that we have developed for the Core expression language. Indeed, our work is only the start. In addition to the passes that we have considered in this paper, we have also translated a few other Core optimization passes, including those for common subexpression elimination and call-arity optimization. Furthermore, we could also prove additional properties about the operations that we have analyzed, such as semantics preservation.

Translating more of GHC would require additional extensions to hs-to-coq. For example, several of the GHC optimization passes are still untranslatable, due to heavy use of mutual recursion. The edits discussed in Section 5 are not sufficient to show the that functions are terminating, yet alternative approaches such as deferred termination [11] are not available. Approaches such as extending hs-to-coq with support for axiomatizing the behavior of mutually recursive functions [5] or integrating it with the Equations package [38] seem promising.

As discussed in Section 5, our edits remove information about rules and unfolding information from Core AST terms. This works well for passes, such as exitify, that do not care about this information, but what about those that do? What if we wanted to reason about this data? Or what if we wanted to reason about other data that is coinductively represented in GHC? One approach would be to augment the hs-to-coq edit language so that it can transform coinductive representations to inductive models, perhaps by storing this information elsewhere in the AST or in additional arguments to compiler passes.

In future work, we would like both to verify more optimization passes and to check more invariants about the Core data structures. In particular, the abstract syntax tree used for GHC's intermediate Core language (Figure 2) is very simple, but this isn't the whole story: GHC maintains and relies on many more invariants of this data structure besides the scoping and join point invariants considered here.

In particular, there are additional structural invariants to reason about, such as that in a case expression, a default case must come first and the other cases must be ordered. More significantly, Core is a typed language, and we could also verify that Core passes preserve the typing of Core terms. The Core type system includes complicated rules like the *let/app invariant*, which govern when the right-hand side of a Let or the argument in an application can be of unlifted types. Furthermore, connecting our manually written invariants to the Core type checker (CoreLint) would be interesting future work.

Finally, we would also like to reason not just about Core invariants but about the semantics for Core terms. It is not difficult to define a simple call-by-name semantics for this AST, which would let us argue that operations are semantics preserving.

In the end, although our efforts are tailored to GHC and its invariants, they reflect the general challenges and benefits of bringing interactive verification to existing large scale software systems, especially those developed using pure functional programming. In particular, our work demonstrates that mechanical reasoning is possible and beneficial even at scale, and even for mature code not developed in conjunction with its proof of correctness.

## Acknowledgments
Thanks to Leonidas Lampropoulos and William Mansky for their valuable comments on a draft of this paper. Thanks to Josh Cohen, Christine Rizkallah, and John Wiegley for assistance with the development of hs-to-coq and the theory of the containers library. This material is based upon work supported by the National Science Foundation under Grant No. 1521539.

## A  Core expression invariants
### A.1  Well-scoped expressions

```
Definition WellScopedVar (v : Var) (in_scope : VarSet) : Prop :=
  if isLocalVar v then
   match lookupVarSet in_scope v with
    | None => False
    | Some v' => almostEqual v v' /\ GoodVar v
    end
  else GoodVar v

Definition GoodLocalVar (v : Var) : Prop :=
  GoodVar v /\ isLocalVar v = true.

Fixpoint WellScoped (e : CoreExpr) (in_scope : VarSet) {struct e} : Prop :=
  match e with
  | Mk_Var v => WellScopedVar v in_scope
  | Lit l => True
  | App e1 e2 => WellScoped e1 in_scope /\  WellScoped e2 in_scope
  | Lam v e => GoodLocalVar v /\ WellScoped e (extendVarSet in_scope v)
  | Let bind body =>
      WellScopedBind bind in_scope /\
      WellScoped body
        (extendVarSetList in_scope (bindersOf bind))
  | Case scrut bndr ty alts  =>
    WellScoped scrut in_scope /\
    GoodLocalVar bndr /\
    Forall' (fun alt =>
      Forall GoodLocalVar (snd (fst alt)) /\
      let in_scope' := extendVarSetList
         in_scope (bndr :: snd (fst alt))
      in WellScoped (snd alt) in_scope') alts
  | Cast e _ =>   WellScoped e in_scope
  | Mk_Type _  =>   True
  | Mk_Coercion _ => True
  end
with WellScopedBind (bind : CoreBind) (in_scope : VarSet) : Prop :=
  match bind with
  | NonRec v rhs =>
    GoodLocalVar v /\
    WellScoped rhs in_scope
  | Rec pairs =>
    Forall (fun p => GoodLocalVar (fst p)) pairs /\
    NoDup (map varUnique (map fst pairs)) /\
    Forall' (fun p => WellScoped (snd p)
      (extendVarSetList in_scope (map fst pairs))) pairs
  end.

Definition WellScopedProgram (pgm : CoreProgram) : Prop :=
   NoDup (map varUnique (bindersOfBinds pgm)) /\
   Forall' (fun p => WellScoped (snd p) (mkVarSet (bindersOfBinds pgm))) (flattenBinds pgm).
```

### A.2  Join Points
The statement of this property is not as elegant as we might wish due to contortions to please the Coq termination checker.

```coq
Definition isJoinPointsValidPair_aux
  isJoinPointsValid isJoinRHS_aux
  (v : CoreBndr) (rhs : CoreExpr) (jps : VarSet) : bool :=
    match isJoinId_maybe v with
    | None => isJoinPointsValid rhs 0 emptyVarSet  (* Non-tail-call position *)
    | Some a =>
      if (a =? 0) (* Uh, all for the termination checker *)
      then isJoinPointsValid rhs 0 jps           (* tail-call position *)
      else isJoinRHS_aux a rhs jps               (* tail-call position *)
    end.

Fixpoint isJoinPointsValid (e : CoreExpr) (n : nat) (jps : VarSet) {struct e} : bool :=
  match e with
  | Mk_Var v => match isJoinId_maybe v with
    | None => true
    | Some a => isLocalVar v && (a <=? n) && elemVarSet v jps
    end
  | Lit l => true
  | App e1 e2 =>
    isJoinPointsValid e1 (n+1) jps &&    (* Tail-call-position *)
    isJoinPointsValid e2 0 emptyVarSet   (* Non-tail-call position *)
  | Lam v e =>
    negb (isJoinId v) &&
    isJoinPointsValid e 0 emptyVarSet    (* Non-tail-call position *)
  | Let (NonRec v rhs) body =>
      isJoinPointsValidPair_aux isJoinPointsValid isJoinRHS_aux v rhs jps &&
      let jps' := updJPS jps v in
      isJoinPointsValid body 0 jps'
  | Let (Rec pairs) body =>
      negb (List.null pairs) &&  (* Not join-point-specific, could be its own invariant *)
      (forallb (fun p => negb (isJoinId (fst p))) pairs ||
       forallb (fun p =>       isJoinId (fst p))  pairs) &&
      let jps' := updJPSs jps (map fst pairs) in
      forallb (fun '(v,e) => isJoinPointsValidPair_aux isJoinPointsValid isJoinRHS_aux v e jps') pairs &&
      isJoinPointsValid body 0 jps'
  | Case scrut bndr ty alts  =>
    negb (isJoinId bndr) &&
    isJoinPointsValid scrut 0 emptyVarSet &&  (* Non-tail-call position *)
    let jps' := delVarSet jps bndr in
    forallb (fun '(dc,pats,rhs) =>
      let jps'' := delVarSetList jps' pats  in
      forallb (fun v => negb (isJoinId v)) pats &&
      isJoinPointsValid rhs 0 jps'') alts  (* Tail-call position *)
  | Cast e _ =>    isJoinPointsValid e 0 jps
  | Mk_Type _  =>   true
  | Mk_Coercion _ => true
  end
with isJoinRHS_aux (a : JoinArity) (rhs : CoreExpr) (jps : VarSet) {struct rhs} : bool :=
  if a <? 1 then false else
  match rhs with
    | Lam v e => negb (isJoinId v) &&
                 if a =? 1
                 then isJoinPointsValid e 0 (delVarSet jps v) (* tail-call position *)
                 else isJoinRHS_aux (a-1) e (delVarSet jps v)
```

```coq
      | _ => false
      end.

Definition isJoinRHS rhs a jps :=
      if (a =? 0)
      then isJoinPointsValid rhs 0 jps
      else isJoinRHS_aux a rhs jps.

Definition isjoinPointsAlt : CoreAlt -> VarSet -> bool :=
  fun '(dc,pats,rhs) jps =>
      let jps'' := delVarSetList jps pats  in
      forallb (fun v => negb (isJoinId v)) pats &&
      isJoinPointsValid rhs 0 jps''.

Definition isJoinPointsValidPair := isJoinPointsValidPair_aux isJoinPointsValid isJoinRHS_aux.

(**
Conjuction of [isJoinId] and [isJoinPointsValidPair]
*)

Definition isValidJoinPointsPair
  (v : CoreBndr) (rhs : CoreExpr) (jps : VarSet) : bool :=
    match isJoinId_maybe v with
    | None => false (* NB *)
    | Some a => isJoinRHS rhs a jps
    end.

(** Join-point validity of whole programs *)

Definition isJoinPointsValidProgram (pgm : CoreProgram)  :=
  Forall (fun '(v,e) =>
    isJoinId v = false /\
    isJoinPointsValid e 0 emptyVarSet = true) (flattenBinds pgm).
```

## B  Axioms

Our proofs of `WellScoped_substExpr` and `exitifyProgram_WellScoped_JPV` rely on the following axioms (listed in Coq via `Print Assumptions`).

**Logic**

These axioms are known to be consistent with Coq's logic.

```coq
FunctionalExtensionality.functional_extensionality_dep :
  forall (A : Type) (B : A -> Type) (f g : forall x : A, B x),
    (forall x : A, f x = g x) -> f = g
ProofIrrelevance.proof_irrelevance : forall (P : Prop) (p1 p2 : P), p1 = p2
JMEq.JMeq_eq : forall (A : Type) (x y : A), x ~= y -> x = y
```

**Deferred fix**

Prior work [11, 12] showed these axioms consistent with Coq. The first is trivially consistent because the return type is inhabited through the `Err.Default` type class. The second is derivable from the axiom of choice (as provided by the Coq module `Coq.Logic.Epsilon`).

```coq
DeferredFix.deferredFix : forall a r : Type, Err.Default r -> ((a -> r) -> a -> r) -> a -> r
```

```
DeferredFix.deferredFix_eq_on :
  forall (a b : Type) (H : Err.Default b)
  (f : (a -> b) -> a -> b) (P : a -> Prop)
  (R : a -> a -> Prop),
  well_founded R ->
  DeferredFix.recurses_on P R f ->
  forall x : a,
  P x ->
  DeferredFix.deferredFix f x = f (DeferredFix.deferredFix f) x
```

**Assertions**

Axioms that stand in for GHC's error reporting mechanism. We do not wish to reason about GHC's error messages, but we do want to know if assertions are violated. Therefore, we do not provide a concrete definition for debugIsOn, forcing verification to consider both alternatives, and we axiomatize the reporting of an assertion failure. All of these axioms can be trivially inhabited, so they do not jeopardize the consistency of Coq.

```
Util.debugIsOn : bool
Panic.someSDoc : String
Panic.assertPanic : forall a : Type, Err.Default a -> String -> Int -> a
Panic.pgmError : forall a : Type, Err.Default a -> String -> a
Panic.panicStr : forall a : Type, Err.Default a -> String -> String -> a
```

**`ValidVarSet`**

The `ValidVarSet` axiom described in Section 6.4.

**Working with `uniqAway` and other axioms related to variables**

The `uniqAway` axiom and its properties described in Section 6.6. We also have an axiom that states that we can tell whether a unique should be used for a local variable (Section 6.5), and an axiom that says that one particular unique used in `exitify` is local unique.

```
Unique.isLocalUnique : Unique.Unique -> bool
Axioms.isLocalUnique_initExitJoinUnique :
    Unique.isLocalUnique Unique.initExitJoinUnique = true
```

**Abstracted parts of the Core language**

Our GHC definition keeps various parts of the Core AST abstract.

```
AxiomatizedTypes.Unbranched : Type
AxiomatizedTypes.Type_ : Type
AxiomatizedTypes.TyThing : Type
AxiomatizedTypes.ThetaType : Type
AxiomatizedTypes.PrimOp : Type
AxiomatizedTypes.ForeignCall : Type
AxiomatizedTypes.DataConBoxer : Type
AxiomatizedTypes.Coercion : Type
AxiomatizedTypes.CoAxiom : Type -> Type
AxiomatizedTypes.CType : Type
AxiomatizedTypes.BuiltInSynFamily : Type
AxiomatizedTypes.Branched : Type
```

We also assume the existence of some operations that work with these types in the AST. However, we do not need to reason about any of the properties of these operations because they do not occur in paths that are considered by our proofs.

```
Core.nopSig : StrictSig
Core.topDmd : Demand
Core.substCo : Util.HasDebugCallStack -> TCvSubst ->
```

```
                  AxiomatizedTypes.Coercion -> AxiomatizedTypes.Coercion
CoreSubst.substTyUnchecked : TCvSubst ->
  AxiomatizedTypes.Type_ -> AxiomatizedTypes.Type_
CoreSubst.substSpec : Subst -> Id -> RuleInfo -> RuleInfo
CoreUtils.exprType : CoreExpr -> AxiomatizedTypes.Type_
```

The only further assumption that we need to make about the above axioms to ensure their consistency is that the `Type_` data structure is inhabited.

```
AxiomatizedTypes.Default__Type_ : Err.Default AxiomatizedTypes.Type_
```